%% file: main.tex
\documentclass[conference]{IEEEtran}
\IEEEoverridecommandlockouts
\usepackage{cite}
\usepackage{amsmath,amssymb,amsfonts}
\usepackage{algorithmic}
\usepackage{graphicx}
\usepackage{textcomp}
\usepackage{xcolor}
\usepackage{subcaption}
\usepackage{mathtools}
\usepackage{url}
\graphicspath{ {./images/} }
\def\BibTeX{{\rm B\kern-.05em{\sc i\kern-.025em b}\kern-.08em
    T\kern-.1667em\lower.7ex\hbox{E}\kern-.125emX}}
\DeclarePairedDelimiter{\ceil}{\lceil}{\rceil}
\DeclarePairedDelimiter\abs{\lvert}{\rvert}
\newcommand{\ourmethod}{\textit{DriCon}}
\newcommand{\micro}{\textit{micro-events}}

\begin{document}

\title{\ourmethod{}: On-device Just-in-Time Context Characterization for Unexpected Driving Events
}
\author{
\IEEEauthorblockN{Debasree Das, Sandip Chakraborty, Bivas Mitra}
\IEEEauthorblockA{Department of Computer Science and Engineering, Indian Institute of Technology Kharagpur, INDIA 721302\\
Email: \{debasreedas1994, sandipchkraborty, bivasmitra\}@gmail.com}}

\maketitle

\begin{abstract}
Driving is a complex task carried out under the influence of diverse spatial objects and their temporal interactions. Therefore, a sudden fluctuation in driving behavior can be due to either a lack of driving skill or the effect of various on-road spatial factors such as pedestrian movements, peer vehicles' actions, etc. Therefore, understanding the context behind a degraded driving behavior just-in-time is necessary to ensure on-road safety. In this paper, we develop a system called \ourmethod{} that exploits the information acquired from a dashboard-mounted edge-device to understand the context in terms of micro-events from a diverse set of on-road spatial factors and in-vehicle driving maneuvers taken. \ourmethod{} uses the live in-house testbed and the largest publicly available driving dataset to generate human interpretable explanations against the unexpected driving events. Also, it provides a better insight with an improved similarity of $80$\% over $50$ hours of driving data than the existing driving behavior characterization techniques. 
\end{abstract}

\maketitle
\input{sections/Introduction}
\input{sections/Related_Work}

\input{sections/Motivation}
\input{sections/System_Overview}
\input{sections/Methodology}
\input{sections/Evaluation}

\input{sections/Conclusion}
\bibliographystyle{IEEEtran}
\bibliography{./bibliography/references}
\newpage


\end{document}

%% file: sections/Introduction.tex
\section{Introduction}\label{intro}
With an increase in the traffic population, we witnessed a phenomenal rise in road accidents in the past few years. According to the World Health Organization (WHO)~\cite{who2022}, the loss is not only limited to humans but affects the GDP of the country as well. The officially reported road crashes are inspected mostly based on the \emph{macro} circumstances, such as the vehicle's speed, the road's situation, etc. Close inspection of those \emph{macro} circumstances reveals a series of \micro{}, which are responsible for such fatalities. For example, suppose a driver hit the road divider and faced an injury while driving on a non-congested road. From the macro perspective, we might presume it is due to the driver's amateurish driving skill or the vehicle's high speed. But, it is also possible that some unexpected obstacles (say, crossing pedestrians/animals) arrived at that moment out of sight. The driver deviated from his lane while decelerating to avoid colliding with them. Therefore, recording these \micro{} are crucial in identifying the reasoning behind such accidents. Such contextual information, or \micro{}, thus, can help various stakeholders like car insurance or app-cab companies to analyze the on-road driving behavior of their drivers. Interestingly, an app-cab company can penalize or incentivize their drivers based on how they handle such context and take counter-measures to avoid accidents.



A naive solution to extract the context information is to analyze the traffic videos. Notably, CCTV cameras~\cite{japan} capture only static snapshots of the events concerning the moving vehicles. Existing works~\cite{japan, nhtsa} use dash-cam videos along with IMU sensor data for manual or partly automated investigation of the accident. Note that, human intervention is error-prone and labor-intensive with higher costs. The situation gets further complicated when multiple events are responsible for the accident. For instance, suppose the preceding vehicle suddenly brakes to avoid collision with a pedestrian or at a run-yellow traffic signal. Consequently, the ego vehicle has to decelerate abruptly, resulting in a two-step chain of responsible events for the unexpected stop. Thus, identifying spatiotemporal interactions among traffic objects are crucial in characterizing the root cause behind such incidents.


\begin{figure}[!t]
    \centering
    \includegraphics[width=0.9\columnwidth,keepaspectratio]{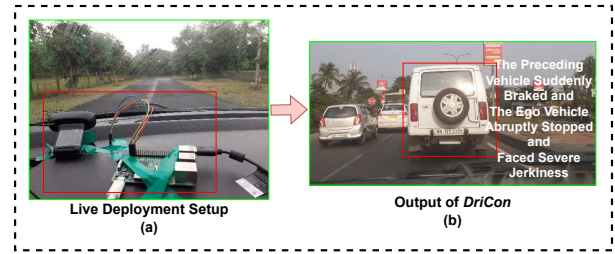}
    \caption{\ourmethod{}: Hardware components and a running instance when a vehicle faced severe jerks}
    \label{fig:edge-out}
\end{figure}
Importantly, understanding the contexts behind the degraded driving behavior on the fly is not trivial and poses multiple challenges. \textbf{First}, this involves continuous monitoring of the driving behavior of the driver as well as an exhaustive knowledge of various on-road spatial \micro{}. Expensive vehicles use LiDAR, Radar etc., to sense the driver and the environment~\cite{lidar,li2021reits}; however, app-cab companies are resistant to invest in such high-end vehicles due to low-profit margin. \textbf{Second}, depending on the driving maneuvers taken, \textit{temporally interlinking the \micro{}} based on the vehicle's interaction with on-road spatial objects is a significant research challenge. For example, if adverse snowy weather is observed on one day, its effect on traffic movements may last till the next day. In contrast, reckless driving would impact only a few other vehicles around and will not be temporally significant after a few minutes. Such temporal impacts of an event would vary depending on the type and space of the event. \textbf{Third}, \emph{spatial positions} of the surrounding objects impact the driving maneuver. Precisely, along with temporal dependency, the distance between the ego vehicle and the surrounding objects plays a vital role. For example, a far-sighted pedestrian might cross the road at high speed, keeping a safe distance, but it is fatal if the distance to the vehicle is low. Existing literature~\cite{akikawa2020smartphone,ridel2019understanding} have attempted to identify risky driving, e.g., vehicle-pedestrian interaction, through IMU and video analysis; however, they fail to capture such temporal scaling or the spatial dependency among surrounding objects. \textbf{Fourth}, identifying the context in real-time over an edge-device (such as a dashcam) is essential for providing a just-in-time feedback. But, deploying such a system for context characterization and analysis from multi-modal data over resource-constrained edge-device is not straightforward.





To address these challenges, we propose \ourmethod{} that develops a smart dash-cam mounted on the vehicle's dashboard to characterize the \micro{} to provide just-in-time contextual feedback to the driver and other stakeholders (like the cab companies). It senses the maneuvers taken by the ego vehicle through IMU and GPS sensors. In addition, a front camera mounted on the device itself, is used to analyze the relationship between various on-road \micro{} and the driving maneuvers taken. This facilitates the system to run in each vehicle in a silo and makes it low-cost and lightweight. Fig.~\ref{fig:edge-out} shows a snapshot of the hardware components of our system mounted on a vehicle, and an example scenario where \ourmethod{} generates a live contextual explanation behind a sudden jerk observed in the vehicle. In summary, our contributions to this paper are as follows.\\

\noindent\textbf{(1) Pilot Study to Motivate Micro-Event Characterization:} We perform a set of pilot studies over the Berkeley Deep Drive (BDD) dataset~\cite{yu2018bdd100k}, the largest public driving dataset available on the Internet (as of \today), to investigate the variations in driving behavior depending on various road types, time of the day, day of the week, etc., and highlight the spatiotemporal \micro{} causing abrupt changes in driving maneuvers.
\\

\noindent\textbf{(2) Designing a Human Explainable Lightweight Causal Model:}
The development of \ourmethod{} relies on the (i) IMU \& GPS data to infer the driving maneuvers, and (ii) object detection model \& perspective transformation~\cite{udacity} to detect the surrounding objects and their actions to capture various spatial \micro{}. Subsequently, we identify the spatiotemporal contexts whenever the driving behavior deteriorates during a trip. Finally, we implement Self Organizing Maps (SOMs), a lightweight but effective causal model to capture the spatiotemporal dependency among features to learn the context and generate human-interpretable explanations.\\

\noindent\textbf{(3) Deployment on the Edge:} We deploy the whole architecture of \ourmethod{} on a Raspberry Pi 3 model, embedded with a front camera, IMU and GPS sensors (Fig.~\ref{fig:edge-out}). For this purpose, we make both the IMU and visual processing of the data lightweight and delay-intolerant. Following this, the pre-trained model generates recommendations based on the ongoing driving trip and makes it efficient to run live for just-in-time causal inferences.\\

\noindent\textbf{(4) Evaluating \ourmethod{} on a Live System Deployment and with BDD Dataset:}
We evaluate \ourmethod{} on our live in-house deployment, as well as on the BDD dataset~\cite{yu2018bdd100k} (over the annotated data~\cite{das2022dribe}), comprising $33$ hours and $17$ hours of driving, respectively. We obtain on average $70\%$ and $80\%$ similarity between the derived and the ground-truth causal features, respectively, with top-$3$ and top-$5$ features returned by the model, in correctly identifying the \micro{} causing a change in the driving behavior. Notably, in most cases, we observe a good causal relationship (in terms of average treatment effect) between the derived features and the observed driving behavior. In addition, we perform different studies of the resource consumption benchmarks on the edge-device to get better insights into the proposed model.

%% file: sections/Related_Work.tex
\section{Related Work}\label{rel_work}
Several works have been proposed in the literature on understanding road traffic and its implications for road fatalities. Early research focused on traffic surveillance-based techniques to prevent road accidents. For instance, National Highway Traffic Safety Administration (NHTSA)~\cite{nhtsa} had recorded statistics about fatal accident cases; TUAT~\cite{japan} has been collecting video records from taxis and drivers' facial images since $2005$ to derive injury instances into several classes along with driving behavior estimation. In India, the source of information behind the causes of traffic injuries is the local traffic police~\cite{indiacrash}. In contrast, works like~\cite{fu2018streetnet, patroumpas2018fly} learn the crime type and aviator mobility pattern just-in-time from street view images and raw trajectory streams, respectively. Apart from harnessing videos and crowd-sourced information, several works~\cite{insight, fan2019gazmon} are done on abnormal driving behavior detection by exploiting IMU and GPS data.
To prevent fatal accidents, authors~\cite{walch2019cooperative, lam2016concise, moosavi2017characterizing} try to alert the drivers whenever risky driving signature is observed, such as lane departure or sudden slow-down indicating congestion. However, they have not looked into the effect of neighboring vehicles or other surrounding factors on various driving maneuvers.

Interaction among the ego vehicle and other obstacles, such as pedestrians, adverse weather in complex city traffic, often affects the vehicle’s motion, consequently affecting the driving behavior. Existing studies~\cite{shi2021predicting} reveal that road category, unsignalized crosswalks, and vehicle speed often lead to a disagreement among pedestrians to cross the road, leading to road fatalities. 
A more detailed study~\cite{samsami2021causal,ramanishka2018toward} focuses on causality analysis for autonomous driving, faces infeasibility in real-time deployment. Moreover, they only use a limited set of driving maneuvers, e.g., speed change only. Particularly, causal inferencing is challenging due to high variance in driving data and spurious correlation~\cite{codevilla2019exploring} between traffic objects and maneuvers.
The existing works limit their study by considering only static road attributes or relying on single or multi-modalities from a connected road network system. Such methodologies will not be applicable for a single vehicle in real-time deployment unless connected to the system. In contrast, leveraging multi-modalities from onboard vehicle sensors can efficiently characterize the continuous and dynamic contexts behind unexpected driving behavior fluctuations. \ourmethod{} develops a system in this direction.

\begin{figure*}[]
    \begin{subfigure}{0.4\columnwidth}
    \includegraphics[width=\linewidth,keepaspectratio]{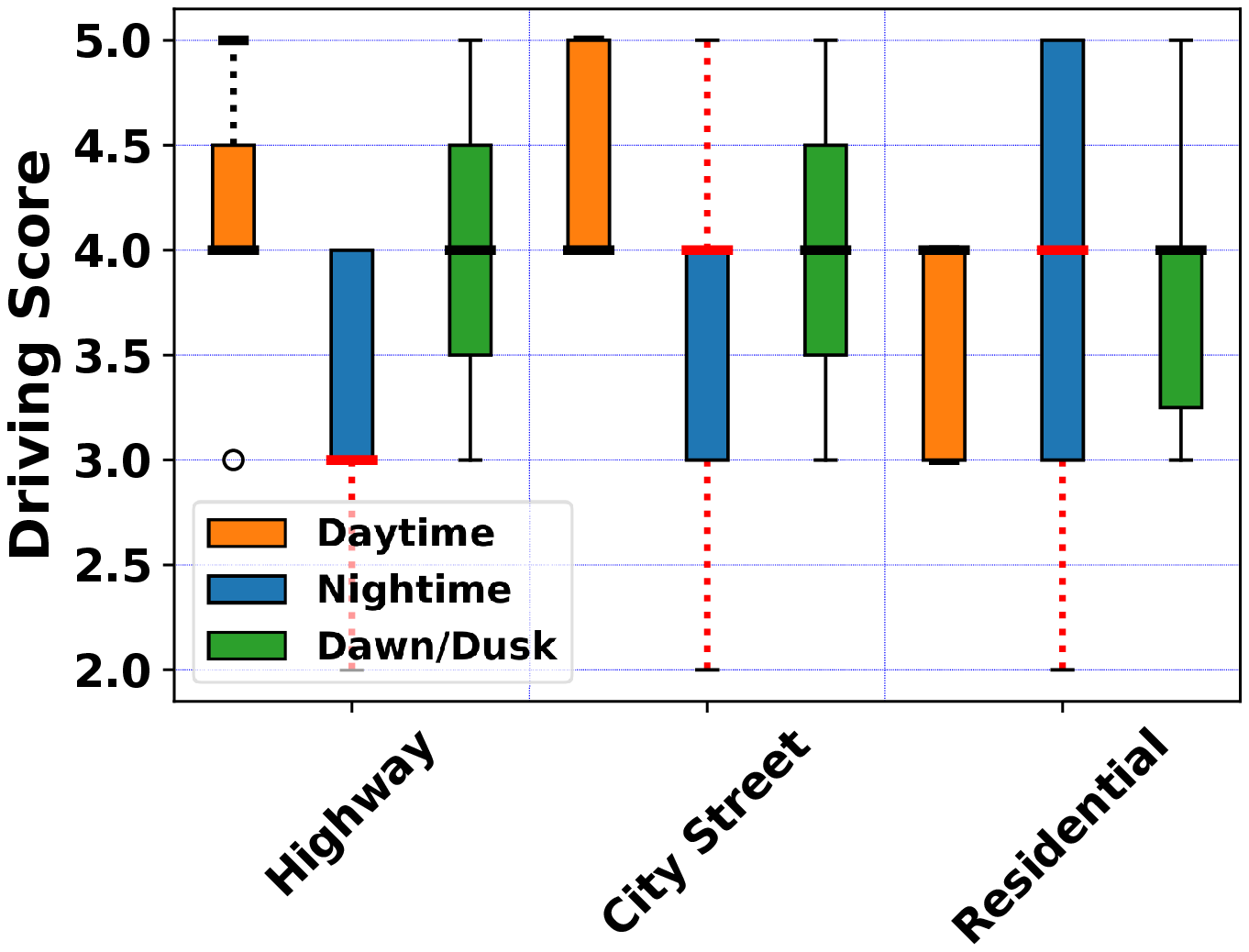}
    \caption{}
    \end{subfigure}
    \begin{subfigure}{0.4\columnwidth}
    \includegraphics[width=\linewidth,keepaspectratio]{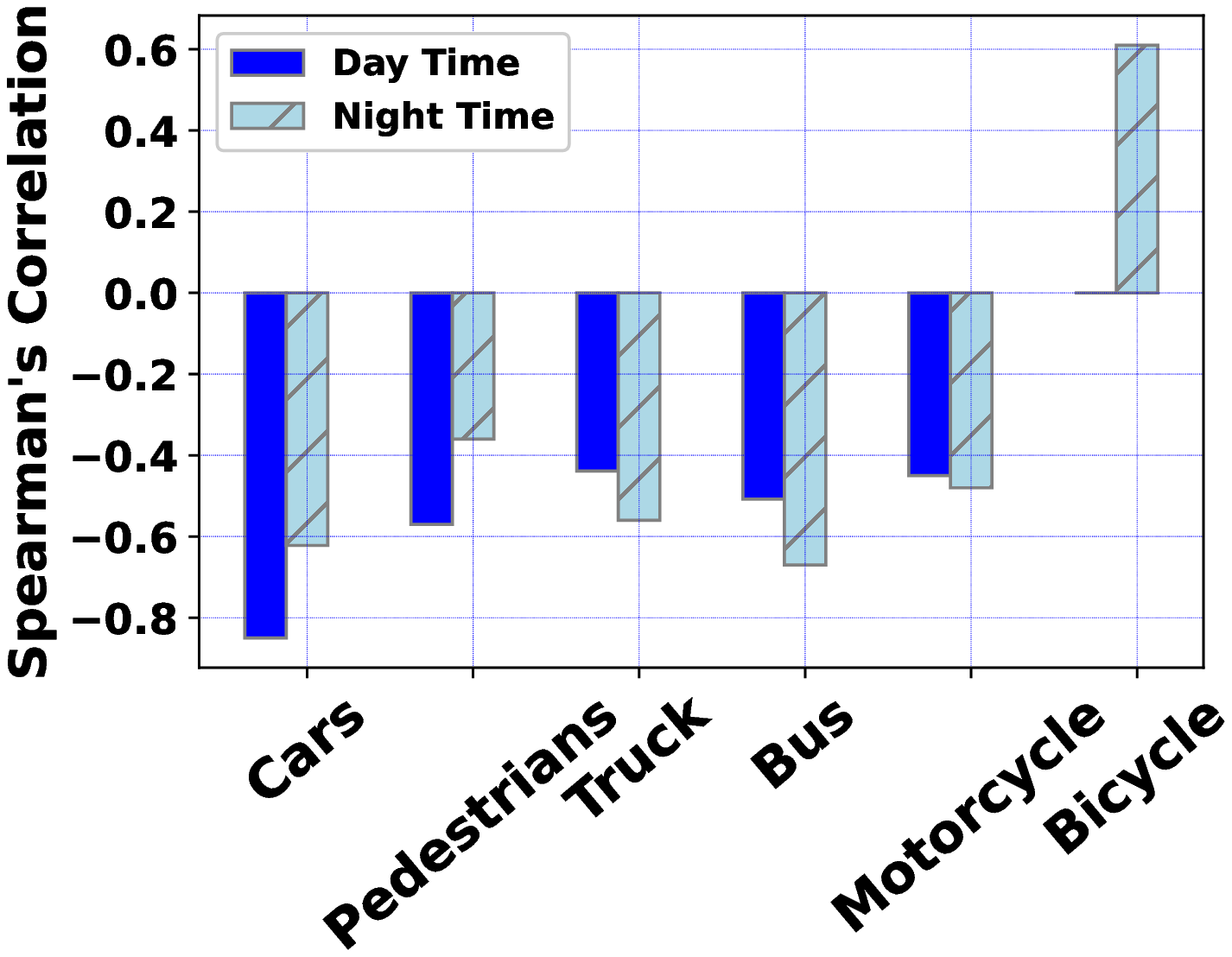}
    \caption{}
    \end{subfigure}
    \begin{subfigure}{0.4\columnwidth}
    \includegraphics[width=\linewidth,keepaspectratio]{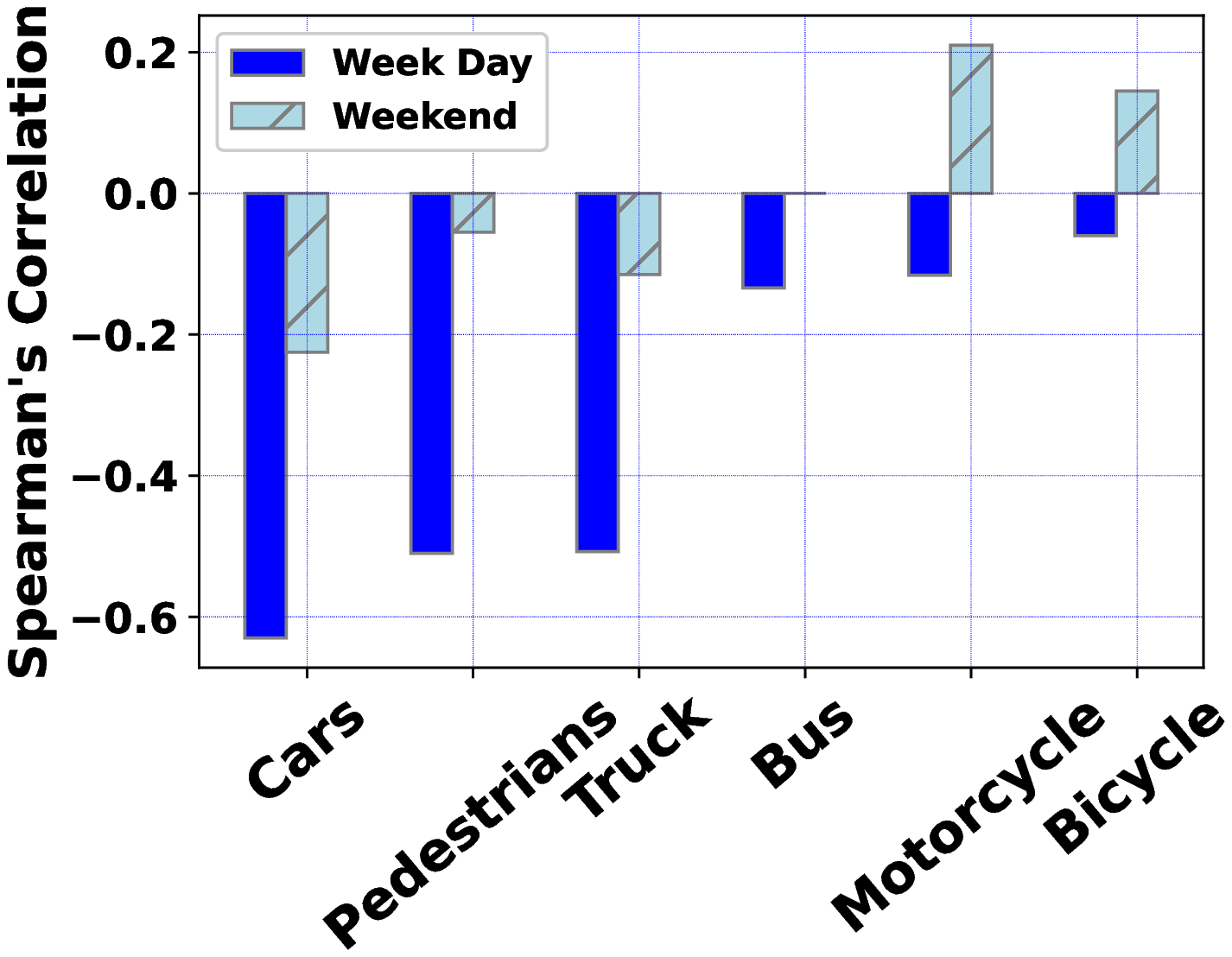}
    \caption{}
    \end{subfigure}
    \begin{subfigure}{0.4\columnwidth}
    \includegraphics[width=\linewidth, keepaspectratio]{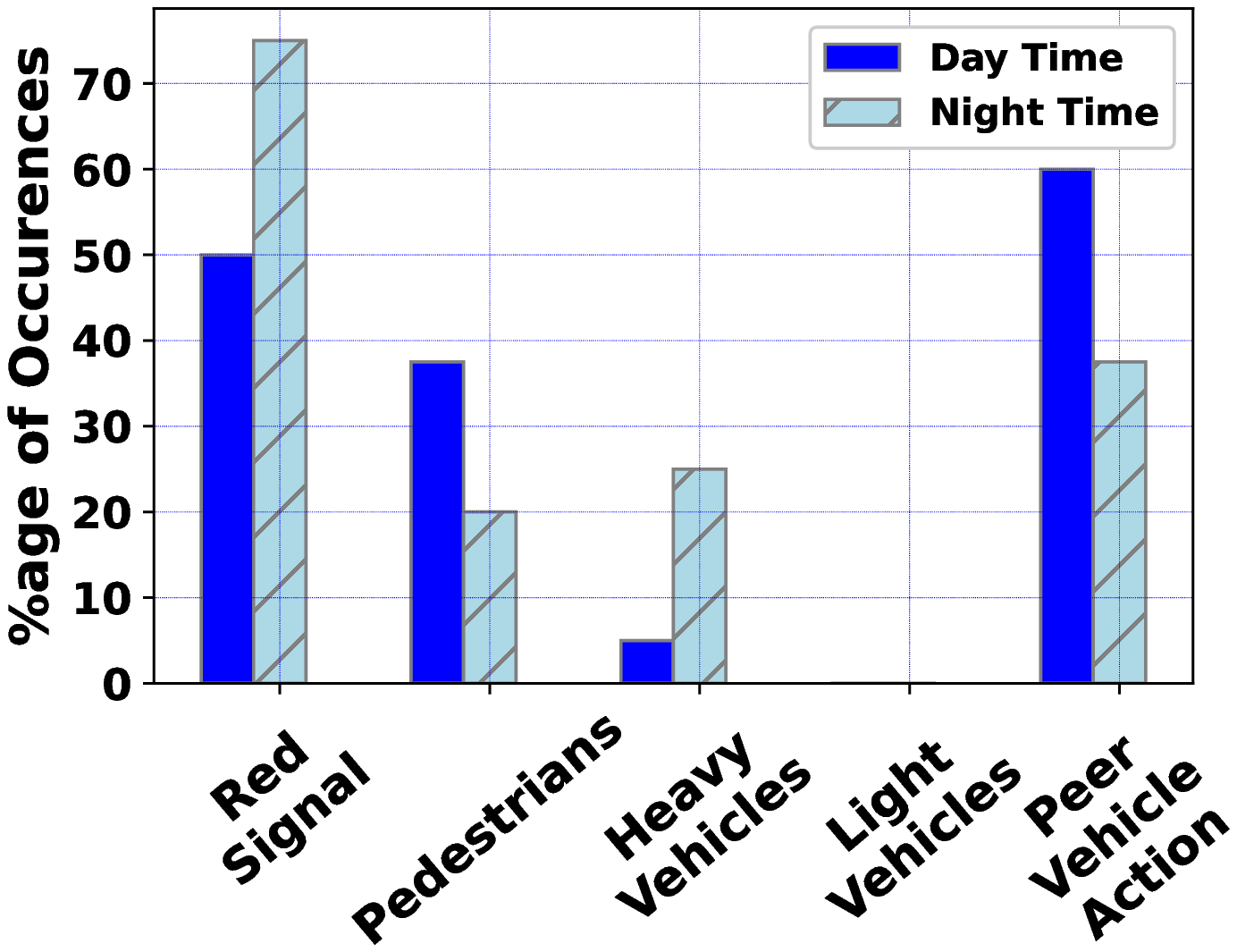}
    \caption{}
    \end{subfigure}
    \begin{subfigure}{0.4\columnwidth}
    \includegraphics[width=\linewidth,keepaspectratio]{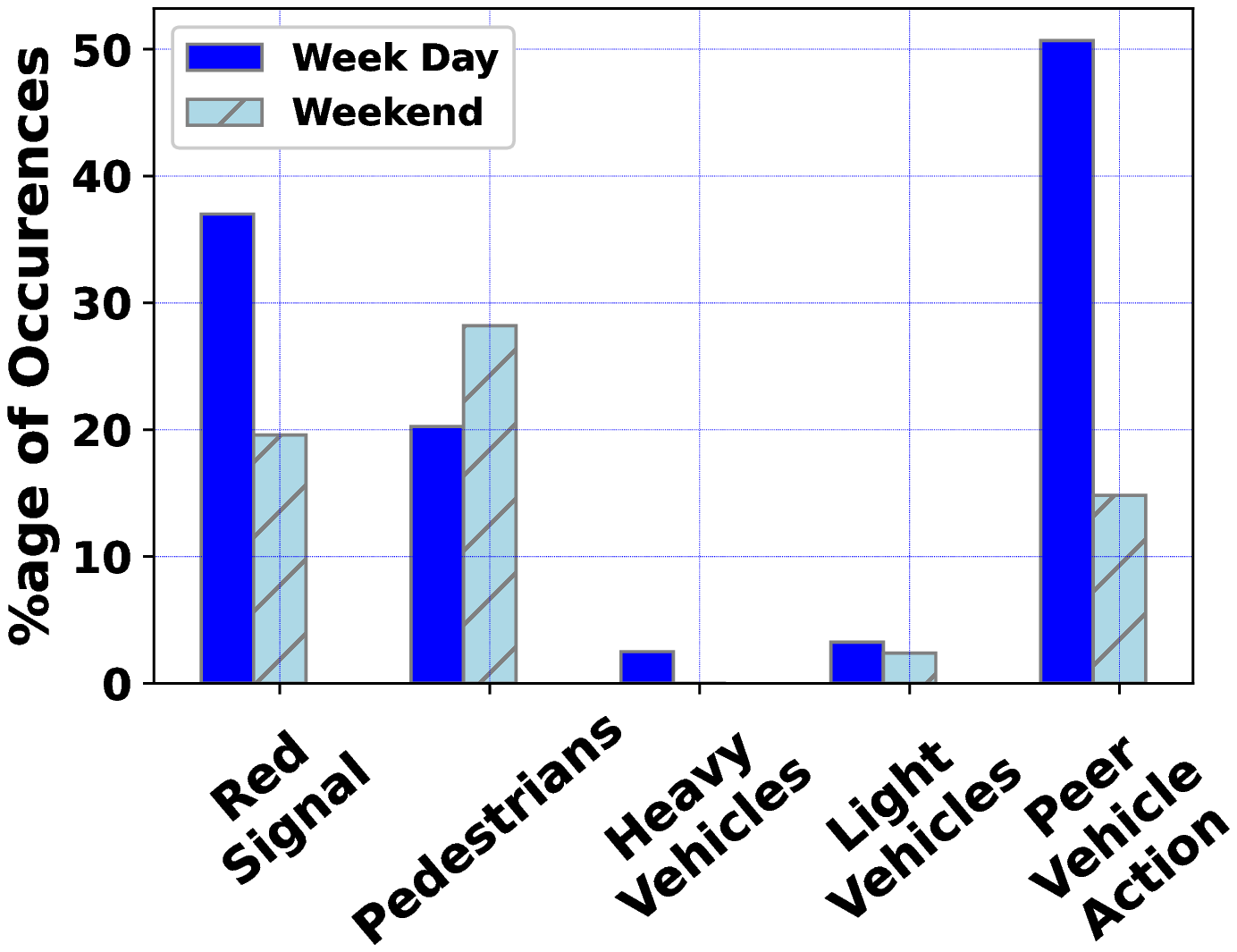}
    \caption{}
    \end{subfigure}
    \caption{(a) Variation of Driving Behavior with respect to Road Type and Time of the Day, (b)-(c) Impact of Spatial \textit{Micro-events} on the Driving Score at Different (i) Time of the Day, (ii) Day of the Week, (d)-(e) Contributing Factors Observed behind Abrupt Stop at Different (i) Time of the Day, (ii) Day of the Week }
    \label{fig:pilot-study}
\end{figure*}

%% file: sections/Motivation.tex
\section{Motivation}\label{motiv}

In an ideal scenario, two vehicles are likely to follow similar maneuvers under the same driving environment; but this is not the case in reality. Driving behavior varies according to the driver's unique skill set and is influenced by the impact of various on-road events, such as the movement of other heavy and light vehicles, movement of pedestrians, road congestion, maneuvers taken by the preceding vehicle, etc., which we call \textit{spatial} \micro{} or \micro{}, in short. In this section, we perform a set of pilot studies to answer the following questions. (a) \textbf{Does a driver's driving behavior exhibit spatiotemporal variations?} (b) \textbf{Do all \micro{} occurrences during a trip similarly impact the driving behavior?} (c) \textbf{Does a sequence of inter-dependent \micro{} collectively influence the driving behavior?} Following this, we analyze the publicly-available open-source driving dataset named Berkeley Deep Drive dataset (BDD)~\cite{yu2018bdd100k} to answer these questions stating the impact of different \micro{} on the driving behavior. The dataset contains $100$k trips crowd-sourced by $10$k voluntary drivers over $18$ cities across two nations -- the USA and Israel. The dataset has been annotated with a driving score on the Likert scale of $1$ (worst driving) to $5$ (best driving) for each $5$-second of driving trips.
\subsection{Variation in Driving Behavior over Space and Time}
We first check whether the on-road driving behavior exhibits a spatiotemporal variation. For this purpose, we vary two parameters -- road type as the \emph{spatial} parameter (say, ``\textit{Highway}'', ``\textit{City Street}'', ``\textit{Residential}''), and time of the day as the \emph{temporal} one (say, ``\textit{Daytime}'', ``\textit{Nighttime}'', ``\textit{Dawn/Dusk}'') in the BDD dataset. In this pilot study, we form $9$ groups with $30$ trips each, in a total of $270$, where the trips under a group are randomly picked from the BDD dataset. We plot the distribution of the  driving scores over all the trips for each group. From Fig.~\ref{fig:pilot-study}(a), it is evident that the score distribution varies both (a) for a single type of road at different times of the day, and (b) for different types of road at any given time of the day (with $p < 0.05$ reflecting its statistical significance). In the following, we investigate the role played by various \micro{} behind the variations in driving behavior.  
\subsection{Role of Spatial \textit{Micro-events}}
Next, we inspect whether various on-road \micro{}, which are characterized by the movements of other spatial objects such as ``\textit{cars}'', ``\textit{pedestrians}'', ``\textit{trucks}'', ``\textit{buses}'', ``\textit{motorcycles}'', ``\textit{bicycles}'', etc., impact a driver's driving behavior in the same way across different times of the day. 
We perform this study by handpicking $30$ trips along with their annotated driving scores for both day and night time from the BDD dataset. We compute the volume (say, count) of spatial objects extracted using the existing object detection algorithm~\cite{yolov3} from the video captured during the trip and take the average count of each object for a $5$-second time window. Thus, for both daytime and nighttime, we get two time-series distributions, (a) the count of each on-road spatial object captured over the trip video during each time window, and (b) the annotated driving scores at those time windows. Next, we compute the Spearman's Correlation Coefficient (SCC) among these two distributions for day time and night time, respectively. From Fig.~\ref{fig:pilot-study}(b), we infer that mostly all the on-road spatial objects adversely affect the driving behavior (depicting a negative correlation). Cars and pedestrians affect the driving score majorly during the daytime. Whereas, at night time, trucks and buses, along with the cars, impact the driving behavior because heavy vehicles such as trucks move primarily during the nighttime. 
However, the effect of light vehicles such as motorcycles and bicycles is insignificant due to the dedicated lanes for their movements. This observation is further extended to Fig.~\ref{fig:pilot-study}(c), where the same study is done for weekdays vs. weekends. We extracted the day of the week using already provided timestamps in the BDD dataset and clubbed $30$ trips from Monday to Friday for weekdays and $30$ trips from Saturday to Sunday for the weekend. From Fig.~\ref{fig:pilot-study}(c), we observe that during the early days of the week, cars, pedestrians, and trucks adversely affect the driving behavior, whereas the impact is less during the weekend. Hence, we conclude that different on-road objects exert diverse temporal effects on the driving behavior. 
\subsection{\textit{Micro-events} Contributing to Sudden Driving Maneuver: Abrupt Stop as a Use-case}
Finally, we explore whether multiple inter-dependent \micro{} can be responsible for a particular driving maneuver that might degrade the driving behavior. For this purpose, we choose \textit{abrupt stop} as the maneuver, which we extract from the GPS and the IMU data (the situations when a stop creates a severe jerkiness~\cite{yu2016fine}). We take $30$ trips for each scenario, including daytime, nighttime, weekdays, and weekends. For each scenario, we extract the instances when an abrupt stop is taken and record the corresponding \micro{} observed at those instances. Precisely, we extract the presence/absence of the following \micro{}: red traffic signal, pedestrian movements, presence of heavy vehicles as {truck \& bus}, light vehicles as {motorcycle \& bicycle}, and the preceding vehicles' braking action (as peer vehicle maneuver), using well-established methodologies~\cite{yolov3, das2022dribe}. We compute the cumulative count of the presence of each \micro{} and the number of abrupt stops taken over all the trips for the four scenarios mentioned above. 
From Fig.~\ref{fig:pilot-study}(d) and (e), we observe that the red traffic signal, the peer vehicle maneuvers, and heavy vehicles mostly cause an abrupt stop during the nighttime and on weekdays.
Therefore, we argue that multiple on-road \micro{}, such as the reckless movement of heavy vehicles at night, force even an excellent driver to slam on the brake and take an unsafe maneuver.

%% file: sections/System_Overview.tex
\section{Problem Statement and System Overview}\label{system}


\subsection{Problem Statement}\label{problem-statement}

Consider that $\mathcal{F}_M$ denotes the set of driving maneuvers and $\mathcal{F}_S$ be the set of spatial \micro{}. $\mathbb{F}^i$ be the set of temporally-represented feature variables corresponding to the driving maneuvers taken and on-road spatial \micro{} encountered during a trip $i$. Let $\mathcal{R}_T^i$ be the driving score at time $T$ during the trip $i$. We are interested in inspecting the events occurred, representing the feature values $\mathbb{F}^i$, when $\abs{\mathcal{R}_T^i - \hat{\mathcal{R}}_{T-1}^i} > \epsilon$ ($\epsilon$ is a hyper-parameter, we set $\epsilon = 1$), reflecting the fluctuations in driving behavior. Here, $\hat{\mathcal{R}}_{T-1}^i = \ceil{\text{mean}([\mathcal{R}_{1}^i, \mathcal{R}_{T-1}^i])}$ represents the mean driving behavior till $T-1$. The output of the system is a characterization of \{$\mathcal{F}_M^i$, $\mathcal{F}_S^i$\}, as to whether a fluctuation in the driving behavior is due to the driving maneuvers only ($\mathcal{F}_M^i$) or forced by the spatially causal \micro{} ($\mathcal{F}_S^i$). Finally, we target to generate the explanations based on \{$\mathcal{F}_M^i$, $\mathcal{F}_S^i$\} to give feedback to the stakeholders for further analysis of the driving profile.

\subsection{Feature Selection}
Leveraging the existing literature~\cite{yu2016fine}, we identified a set of feature variables at timestamp $T$ representing various driving maneuvers $\mathcal{F}_M$ of the ego vehicle. These features are -- Weaving ($\mathcal{A}_T^W$), Swerving ($\mathcal{A}_T^S$), Side-slipping ($\mathcal{A}_T^L$), Abrupt Stop ($\mathcal{A}_T^Q$), Sharp Turns ($\mathcal{A}_T^U$), and Severe Jerkiness ($\mathcal{A}_T^J$). Similarly, we consider the following feature variables corresponding to the spatial \micro{} $\mathcal{F}_S$ -- Relative Speed ($\mathcal{S}_T$) and Distance ($\mathcal{D}_T$) between the ego and the preceding vehicle, preceding vehicle's Braking Action ($\mathcal{B}_T$), volume of the peer vehicles in front of the ego vehicle indicating Congestion in the road ($\mathcal{C}_T$), Pedestrian ($\mathcal{P}_T$), and it's speed ($\mathcal{Q}_T$), Traffic light ($\mathcal{L}_T$), Heavy vehicles: \{Bus \& Truck\} ($\mathcal{H}_T$), Type of  the Road ($\mathcal{G}_T$), and Weather condition ($\mathcal{W}_T$). Note that, we empirically select these features based on the existing literature and observations from the dataset; additional features can also be incorporated in \ourmethod{} without losing its generality.

\begin{figure}[!ht]
    \centering
    \includegraphics[width=0.99\columnwidth,keepaspectratio]{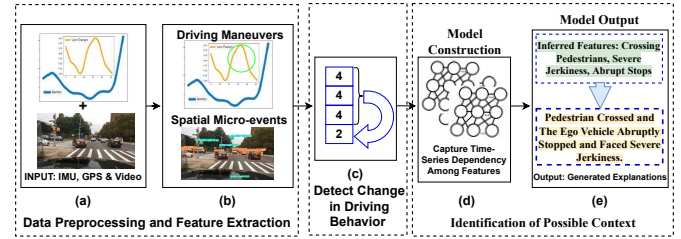}
    \caption{\ourmethod{} System Flow and Modeling Pipeline}
    \label{fig:method-pipe}
\end{figure}

We next broadly introduce our system architecture. \ourmethod{} captures IMU, GPS, and video data from a dashcam (say, an edge-device) and characterizes the context behind the improved/degraded driving behavior on the fly. The system comprises three components: (a) \textbf{Data Preprocessing and Feature Extraction}, (b) \textbf{Detection of Improved/Degraded Driving Behavior}, and (c) \textbf{Identification of Possible Context} (see Fig.~\ref{fig:method-pipe}).

\subsection{Data Preprocessing and Feature Extraction}\label{feature-process}
The collected IMU and GPS sensor data are prone to noise due to the earth's gravitational force, signal attenuation, and atmospheric interference. Hence, we implement a low-pass filter to eliminate such noises from IMU and GPS to compute inertial features for the extraction of the driving maneuvers ($\mathcal{F}_{M}$). Next, we preprocess the video data before extracting on-road spatial \micro{} and their actions ($\mathcal{F}_S$). We up/downsample the acquired videos to a resolution of $960\times540$p, preserving the signal-to-noise ratio above $20$ dB.


\subsubsection{\textbf{Driving Maneuvers - $\mathcal{F}_M$}}
In order to generate the features corresponding to different driving maneuvers ($\mathcal{F}_{M}$), we extract the instances of {Weaving ($\mathcal{A}_T^W$), Swerving ($\mathcal{A}_T^S$), Side-slipping ($\mathcal{A}_T^L$), Abrupt Stop ($\mathcal{A}_T^Q$), Sharp Turns ($\mathcal{A}_T^U$), and Severe Jerkiness ($\mathcal{A}_T^J$)} from the IMU data using standard accelerometry analysis~\cite{yu2016fine, das2022dribe}.
\subsubsection{\textbf{Spatial \textit{Micro-events} - $\mathcal{F}_S$}}\label{traffic-obstacles}
Next, we implement the state-of-the-art video data-based object detection algorithms and further fine-tune them based on our requirements, as developing vision-based algorithms is beyond the scope of our work. We leverage the YOLO-V3~\cite{yolov3} algorithm trained on the COCO dataset~\cite{lin2014microsoft} to detect a subset of traffic objects such as \textbf{Pedestrians}, \textbf{Cars}, \textbf{Buses}, \textbf{Trucks}, and \textbf{Traffic Lights} (depicted as $\mathcal{F}_{S}$). Next, we estimate the influence of pedestrians' interactions, the presence of heavy vehicles~(buses \& trucks), traffic light signal transitions (red, yellow \& green), and the cars on the driving behavior of the ego vehicle. Next, we discard the detected objects which depict a confidence score $<50\%$ and bounding boxes of area $<10k$, capturing the fact that the far-sighted traffic objects around the ego vehicle exert marginal impact compared to the near-sighted ones. Additionally, the traffic objects in the mid-way of the road, broadly visible from the driver's dashboard, will be of more influence than the left or right lanes, as the ego vehicle will follow them immediately. Thus, we divide each of the frames into $0.2$:$0.6$:$0.2$ ratio along the horizontal axis, as left:middle:right lanes.
Therefore, we keep the \textbf{Pedestrians} $\mathcal{P}_T$, \textbf{Cars}, \textbf{Heavy Vehicles} as \{\textbf{Buses} \& \textbf{Trucks}\}  $\mathcal{H}_T$, which have bounding box co-ordinates within the middle lane boundary, and \textbf{Traffic Light Signal Transitions} $\mathcal{L}_T$ (Red, Yellow \& Green) without the lane information as traffic lights are often positioned on the left and right lanes.
Since our pilot study demonstrated that the pedestrians and peer vehicles' action significantly impact the driving maneuvers of the ego vehicle, (a) we extract the \textbf{Pedestrian Speed} ($\mathcal{Q}_T$), as well as identify the crossing pedestrians in the mid-way, and (b) we compute the preceding vehicle's \textbf{Braking Action ($\mathcal{B}_T$)}, and \textbf{Congestion ($\mathcal{C}_T$)}, as well as detect the \textbf{Relative Speed ($\mathcal{S}_T$) and Distance ($\mathcal{D}_T$)} variation among the ego and the preceding vehicle.
We apply perspective transformation and deep learning methods~\cite{udacity,das2022slammed} to infer the above. Finally, the above pipeline runs on each frame where the video is re-sampled to $15$ frames-per-second.




\subsection{Detection of Driving Behavior Fluctuations}\label{score-fluct}
The crux of \ourmethod{} is to capture the temporal dependency of various driving maneuvers and spatial \micro{} when a change in the driving behavior is observed during the trip. For a run-time annotation of the driving behavior, we use an existing study~\cite{das2022dribe} that provides a driving behavior score on the Likert scale $[1-5]$ by analyzing driving maneuvers and other surrounding factors. We divide the trip into continuous non-overlapping time windows of size $\delta$ and compute the driving score at the end of every window $\mathcal{U}$ (denoted as $\mathcal{R}^P_{\mathcal{U}}$), using the feature values captured during that window~\cite{das2022dribe}. To quantitatively monitor whether there is a change in the driving behavior during a window $\mathcal{U}$, we compare $\mathcal{R}^P_{\mathcal{U}}$ and $\hat{\mathcal{R}}^P_{\mathcal{U}} = \frac{1}{\mathcal{U}-1}\sum\limits_{i=1}^{\mathcal{U}-1}\mathcal{R}^P_i$ (mean driving score during previous $\mathcal{U}-1$ windows). Suppose this difference is significant (greater than some predefined threshold $\epsilon$). In that case, \ourmethod{} proceeds towards analyzing the temporal dependency among the feature vectors at different time windows to understand the reason behind this difference.


\subsection{Identification of Possible Context}
In the final module, we use the feature vectors at different windows to build the model that identifies which features ($\mathcal{F}_{GEN}$) are responsible for the change in driving behavior during the window $\mathcal{U}$. The model reactively seeks explanations behind such fluctuations by analyzing the effect of the \micro{} that occurred over the past windows $[1,\cdots,(\mathcal{U}-1)]$ and the present window $\mathcal{U}$. Finally, natural language-based human interpretable explanations are generated and fed back to the stakeholders for further analysis.

%% file: sections/Methodology.tex
\section{Model Development}\label{method}
To develop the core model for \ourmethod{}, we leverage the already extracted features $\mathcal{F} \in \{\mathcal{F}_M \bigcup \mathcal{F}_S\}$ (details in~\S\ref{feature-process}) to capture the temporal dependency of the past as well as the present events. In addition, \ourmethod{} derives the explanation behind the detected events through explanatory features $\mathcal{F}_{GEN}$. For this purpose, we need a self-explanatory model that can capture the spatiotemporal dependency among different driving maneuvers and \micro{} associated with the on-road driving behavior. We choose a \textit{Self Organizing Map} (SOM)~\cite{kohonen1990self} for constructing the model that can exploit such spatiotemporal dependencies with minimum data availability. The major limitation of the classical deep learning models (such as CNN or RNN) stems from the fact that, (i) deep networks consume heavy resources (say, memory), as well as suffer from huge data dependency, and (ii) they act as a black box, hence fail to generate human interpretable explanations behind certain predictions~\cite{rudin2019stop}. On the other hand, SOM is able to characterize the \micro{} in runtime using feature variability and unlabelled data. 


\begin{figure}[!ht]
    \centering
    \includegraphics[width=0.75\columnwidth,keepaspectratio]{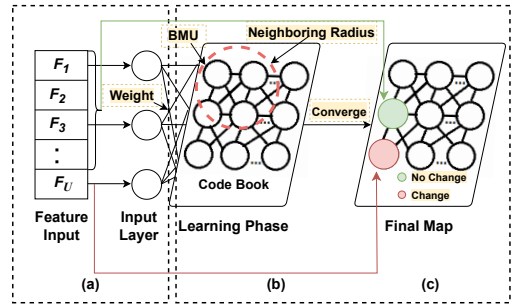}
    \caption{Working Principle of SOM}
    \label{fig:som-arch}
\end{figure}

\subsection{Inferring Explanatory Features using SOM}\label{section:som}
The key idea behind obtaining the explanatory features is first to discover the spatiotemporal feature dependency. In \ourmethod{}, we derive so using Kohonen’s Self Organizing Map (see Fig.~\ref{fig:som-arch}), as it is an unsupervised ANN-based technique leveraging competitive learning methods. Since \ourmethod{} runs on an edge-device, we employ a minimal number of model parameters to expedite the processing without compromising the performance.
Precisely, we implement the \textit{codebook} with $147$ neurons, spread out over a two-dimensional array of size $7\times21$ (where $7$ is a hyperparameter depending on the maximum influence of the past windows during a trip, $21$ corresponds to the number of features in the feature space). These neurons are initialized with a random weight (see Fig.~\ref{fig:som-arch}(a)), where the weight vector has the same length (of $21$) as the feature vector. Next, we represent each trip with a 2D grid of size $8\times21$ (considering $8$ consecutive windows in a trip) to capture the influence of the past windows $[1,\cdots,(\mathcal{U}-1)]$ and the present window $\mathcal{U}$. In principle, the inherent \emph{topological ordering} of SOM groups the similar feature space (in windows $[1,\cdots,(\mathcal{U}-1)]$) into a \emph{single group}, when there is no change in the driving behavior. On the contrary, the dissimilar ones (say, during the window $\mathcal{U}$), when there exists a change in the driving behavior, are mapped into a \emph{different group}, as depicted in Fig.~\ref{fig:som-arch}(b,c).

For instance, suppose on a trip, the ego vehicle abruptly stops due to the preceding vehicle’s braking action following a sudden change in the traffic signal. Hence the feature space in window $[1,\cdots,(\mathcal{U}-1)]$ exhibits a similar signature (until the abrupt stop occurs), and subsequently gets mapped to a \emph{single neuron}. However, during the abrupt stop, there will be changes in the feature space (say, maneuvers and other spatial events). These changes in the feature space will get it assigned to a \emph{different neuron} and settle the other neurons' weight automatically depending on the changes in the feature space between the windows $[1,\cdots,(\mathcal{U}-1)]$ and the window $\mathcal{U}$. This procedure allows SOM to harness the temporal dependency among spatial events in an unsupervised mode, without using the driving score explicitly.



\subsubsection{\textbf{Model Training}} The input trip data is represented in the $2D$ grid format for learning the best-matched neuron, optimizing the Euclidean distance between the feature space and weight vector of the corresponding neuron. To ensure the best-fitting, the best-matched neuron tries to learn the weight vector of the feature space at most. Also, the neurons in the neighborhood try to tune their weights as nearest as possible compared to the best-matched neuron. We train this model for $500$ epochs, where each neuron gets mapped with the best matching trip instances and converges to their coordinate position in the \emph{codebook}. We implement the Bubble neighborhood function~\cite{bubbleneighborhood} to update the neighborhood neurons' weights until the neighborhood radius converges to $\approx 0$. We ensure that both the distance and neighborhood functions are computationally faster for accurate learning accelerating the convergence. Upon completing the total number of epochs, we obtain the converged codebook called the \emph{Map}, where each trip instance gets assigned to the best matching neuron called the \emph{Best Matching Unit (BMU)}. The weight vector corresponding to the BMU's coordinate reveals the explanatory features $\mathcal{F}_{GEN}$.


\subsubsection{\textbf{Model Execution}}
We leverage the constructed \emph{Map} for the runtime inference. First, we conduct the feature processing of the current ongoing trip (following ~\S\ref{feature-process}), and in parallel, the extracted feature space is fed as input to the constructed \emph{Map}. Eventually, we obtain the BMU’s coordinate and extract its corresponding weight vector and the feature encoding for the given trip instance. From the weight vector, we extract the top-k weights and their corresponding feature names (say, \textit{weather type}) and their encoded values (say, \textit{weather type: rainy}). Finally, we populate them in $\mathcal{F}_{GEN}$ (called the \textit{Generative \micro{}}) for further generation of human interpretable explanation.


\subsection{Generating Textual Explanation}\label{section:text}
\ourmethod{} aims to generate the explanations in textual format utilizing the output features $\mathcal{F}_{GEN}$ for better readability and human interpretation. As the features $f \in \mathcal{F}_{GEN}$ are already associated with some keywords (say, \textit{severe jerkiness}), we need to generate them in a sentential form, keeping the features as ``action'' or ``subject'' depending on whether $f \in \mathcal{F}_M$ or $f \in \mathcal{F}_S$, respectively. For instance, if the feature is an \emph{action}, we assign the ego vehicle as the subject, replace the 
corresponding output feature $f$ with its describing keyword, and finally concatenate them to obtain the sentential form. For example, in case of \emph{severe jerkiness}, the constructed sentence becomes, ``\textit{the ego vehicle severe jerks}''. However, if the output feature $f$ represents a \emph{subject}, then many possible sentences can be generated out of one subject. Thus, we mine several traffic guidelines~\cite{guide-pedestrian} and compute the cosine similarity among the features and existing guidelines using TF-IDF vectorizer. Upon extracting the most relevant guidelines, we fetch the object associated with the sentence and construct a single sentence for each output feature (e.g., ``\textit{pedestrian crossing}'' $\rightarrow$ ``\textit{pedestrian crossing the intersection}''). Next, for all the generated sentences, the describing keywords corresponding to each feature are converted to an adjective or adverb using Glove~\cite{pennington2014glove} for better structuring of the sentences (say, ``\textit{the ego vehicle \underline{severe} jerks}'' $\rightarrow$ ``\textit{the ego vehicle \underline{severely} jerks}''). Finally, each sentence is concatenated using the ``\textit{and}'' conjunction, and repetitive subjects are replaced using their pronoun form using string manipulation to generate the whole explanation, as depicted in Fig.~\ref{fig:method-pipe}(e). 

%% file: sections/Evaluation.tex
\section{Performance Evaluation}\label{section:eval}
This section gives the details of \ourmethod{} implemented over a live setup as well as over the BDD dataset. We report the performance of the SOM model and compare it against a well-established baseline. Additionally, we show how well our system has generated the textual explanations along with a sensitivity analysis to distinguish how error-prone \ourmethod{} is. We start with the experimental setup details as follows. 
\subsection{Experimental Setup}\label{exp-setup}
\ourmethod{} is implemented over a Raspberry Pi 3 Model B microprocessor kit operating Raspbian OS with Linux kernel version $5.15.65-v7+$ along with $1$ GB primary memory and ARMv$7$ processor. We primarily utilize the IMU, the GPS, and the video data captured through the front camera (facing towards the front windscreen) as different modalities. For this purpose, we embed one MPU$-9250$ IMU sensor, one u-blox NEO$-6M$ GPS module, and one Logitech USB camera over the Raspberry Pi board, as depicted in Fig.~\ref{fig:edge-out}(a). We deployed \ourmethod{} over three different types of vehicles (e.g., SUV, Sedan, \& Hatchback). We hired $6$ different drivers in the age group of $[20-45]$ who regularly drive in practice. Therefore, our whole experimentation ran for more than two months over three cities, resulting in approximately $33$ hours of driving over $1000$ km distance. The drivers drove freely without any specific instructions given, with each trip varying from approximately $20$ minutes to $2$ hours. In addition, each driver drove over five different types of roads (city street, highway, residential, parking \& campus road) at three different times of the day (day, dusk \& night). We evaluate \ourmethod{} by analyzing how well our proposed model extracts the generative  \micro{} $\mathcal{F}_{GEN}$ (see \S\ref{section:text}). For implementing \ourmethod{}, we consider $\delta = 5$ seconds, $\epsilon = 1$. The impact of other hyperparameters and resource consumption have been discussed later during the analysis. We next discuss the ground-truth annotation procedure used for the evaluation of \ourmethod.

\subsection{Annotating Micro-events}\label{annotate-micro}
We launched an annotation drive by floating a Google form among a set of recruited annotators, where they had to watch a video of at most $10$ seconds and choose the top-$3$ most influential factors impacting the driving behavior. We do this annotation over the in-house data (video data collected during the live experiments) and the videos over the BDD dataset. For each video from both the datasets given in the form, we showed only the clipped portion where the score fluctuations had occurred. Next, out of the total $15$ factors (including driving maneuvers and spatial \micro{}) given in a list, they were instructed to choose the top-$3$ most influential factors responsible for the poor driving behavior based on their visual perception. Besides, we also provided the model-generated sentences (\S\ref{section:text}) and asked how relevant and well-structured the sentences are (on a scale of $[1-5]$) for explaining the change in the driving behavior. The annotators also had the option to write their own explanation if they perceived a better reason behind the driving behavior change. As the number of trips is quite large, we need to design a set of Google forms (sample form\footnote{\url{https://forms.gle/97N6uk4ujRaZSWbj8} (Accessed: \today)}), each containing at most $20$ videos to ensure the least cognitive load on the annotators. We also collected annotators' demographic information such as age, gender, city, etc. We find that most participants ($>67$\%) had prior driving skills. At least three independent annotators had annotated each instance. Upon receiving the annotated factors, we need to find the agreement among the annotators to ensure the received ground truth is unbiased and non-random. As standard inter-annotator agreement policies (say, Cohen's kappa index) work on quantitative analysis or one-to-one mapping, we cannot apply such metrics. Thus, we use the majority voting technique where each listed factor is assigned a percentage, signifying how many times the annotators choose that factor. Each factor having a vote of at least $60$\% is kept in $\mathcal{F}_{GT}$. We observe the minimum and the maximum cardinality of $\mathcal{F}_{GT}$ are $3$ and $5$, respectively. This also indicates that the annotators agreed on selecting the factors that influenced the driving behavior. $\mathcal{F}_{GT}$ contains the annotated \micro{} against which $\mathcal{F}_{GEN}$ is evaluated.
\begin{table*}[]
\centering
\scriptsize
\caption{Similarity Measure among Human Annotated vs. Model Generated Output}
\label{table:eval}
\begin{tabular}{c||c||c||c||l}
\hline\hline
\multicolumn{1}{l||}{\textbf{Instance\#}} & \textbf{Human Annotated $\mathcal{F}_{GT}$}                                                                                                                                 & \textbf{Model Generated $\mathcal{F}_{GEN}$}                                                                                                    & \textbf{Similarity $\mathcal{N}$(\%)} & \textbf{ATE} \\ \hline\hline
\textbf{1}                                & \begin{tabular}[c]{@{}c@{}}Poor Weather Conditions (Heavy Rainfall, Fog, etc.), Swerving,\\  \textbf{Congestion}, Overtaking, \textbf{Taking Abrupt Stop}\end{tabular} & \begin{tabular}[c]{@{}c@{}}\textbf{Congestion}, Preceding Vehicle Braking,\\  Weaving, \textbf{Abrupt Stop}, Severe Jerkiness\end{tabular} & $40$\%                 & $1.96$         \\ \hline
\textbf{2}                                & Sideslip, \textbf{Taking Abrupt Stop}, \textbf{Traffic Lights: Red}                                                                                                    & \textbf{Traffic Lights: Red}, Congestion, \textbf{Abrupt Stop}                                                                             & $66.67$\%                 & $2.5$          \\ \hline
\textbf{3}                                & \textbf{Crossing Pedestrian}, High Speed Variation among Cars, \textbf{Weaving}                                                                                        & Severe Jerkiness, \textbf{Crossing Pedestrian}, \textbf{Weaving}                                                                           & $66.67$\%                 & $1.35$         \\ \hline\hline
\end{tabular}
\end{table*}

\subsection{Performance Metric}
We use the \textbf{Dice Similarity Coefficient score}~\cite{carass2020evaluating} ($\mathcal{N}$) which computes the similarity between $\mathcal{F}_{GT}$ and $\mathcal{F}_{GEN}$ as follows: $\mathcal{N} = \frac{2\times |\mathcal{F}_{GT} \cap \mathcal{F}_{GEN}|}{|\mathcal{F}_{GT}|+|\mathcal{F}_{GEN}|}$. We report the mean $\mathcal{N}$ across all the trips to measure the accuracy of \ourmethod. Next, we also use \textbf{Average Treatment Effect}~\cite{rubin1974estimating} (ATE) to report comparatively higher causal features out of the model identified features. Finally, we define \textbf{Percentage of Error} as follows. First, we compute the set-difference as \{$\mathcal{F}_{GT} \setminus \mathcal{F}_{GEN}$\}, and extract the corresponding feature category (say, $\mathcal{F}_{M}$, $\mathcal{F}_{S}$). Once we get the count of each feature category, we compute its percentage out of the total trips as the \textbf{Percentage of Error}.

\subsection{Baseline Implementation}
As a baseline for extracting $\mathcal{F}_{GEN}$, we implement a supervised rule-based Random Forest (RF) algorithm with $20$ decision trees where each tree is expanded to an unlimited depth over the training data. We optimize the labels $\mathcal{R}^P_\mathcal{U}$ with the intuition that features will contribute differently to each of the predicted scores. Although the RF-based model has a feature importance score signifying the contribution of each feature in constructing the model, we need to have an explanation of how each feature contributes to predicting the driving scores on a trip instance basis. Therefore, we use LIME~\cite{ribeiro2016should} in the background of the RF model for generating the explanatory features. As LIME is a model-agnostic method, it tries to map the relationship between the input features and output scores by tweaking the feature values. Thus, it explains the range of values and probability for each feature that contributes to predicting the score. From the generated explanation, we extract the contributing features $\mathcal{F}_{GEN}$ along with their values for further generation of textual explanation. This pipeline is executed in a similar manner as described in~\S\ref{exp-setup}.


\begin{figure}[]
\centering
\begin{subfigure}{0.45\columnwidth}
    \includegraphics[width=\linewidth,keepaspectratio]{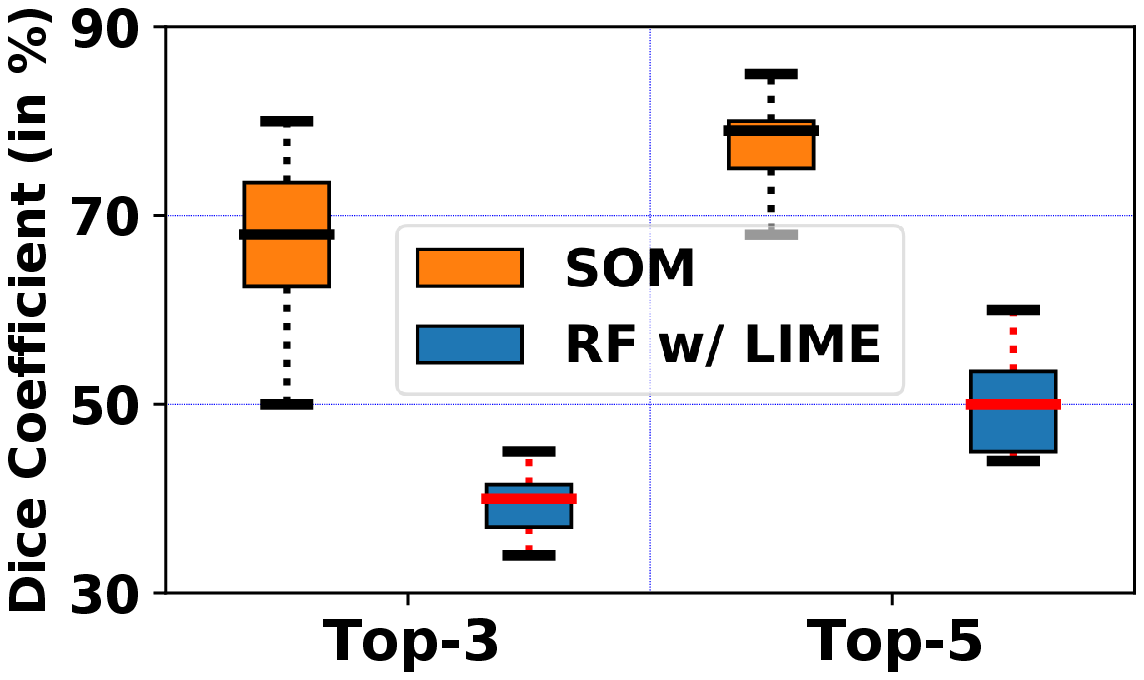}
    \caption{}
    \end{subfigure}
    \begin{subfigure}{0.45\columnwidth}
    \includegraphics[width=\linewidth,keepaspectratio]{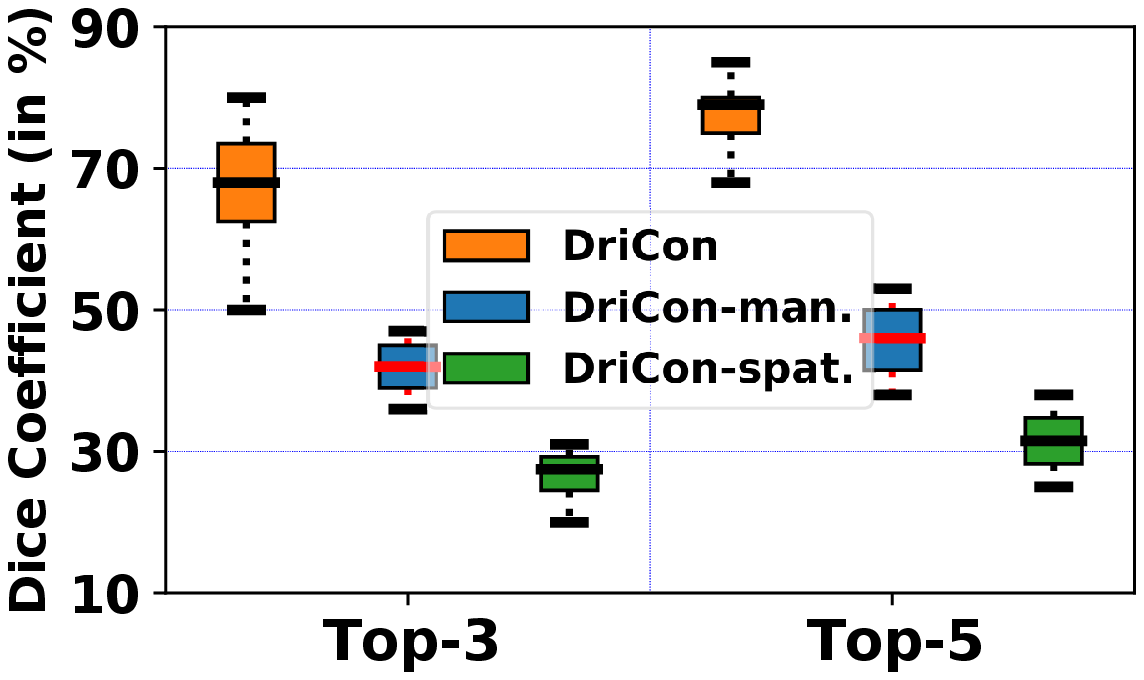}
    \caption{}
    \end{subfigure}
\caption{(a) Dice Coefficient Similarity (in \%) between Human Annotated and Model Generated Features (b) Ablation Study}
\label{eval}
\end{figure}
\subsection{Accuracy of Characterized Context}
We present the accuracy of \ourmethod{} using the SOM and RF+LIME model over the in-house dataset using Dice Coefficient Similarity $\mathcal{N}$. We extract the top-k features from $\mathcal{F}_{GEN}$ where $k \in \{3,5\}$ and compute $\mathcal{N}$ between the two sets of features -- $\mathcal{F}_{GEN}$ and $\mathcal{F}_{GT}$ with top-k. Fig.~\ref{eval}(a) shows the result. For top-3, we get $69\%$ \& $40\%$ similarity on average with SOM and RF+LIME, respectively. Whereas for top-5, we observe $79\%$ \& $48$\% similarity on average with SOM and RF+LIME, respectively.
As the in-house dataset has more complex \micro{}, the slight performance drop over the in-house dataset using the top-3 features is tolerable. Intuitively, the model can capture more diversity as perceived by the human annotators; therefore, the similarity improves as we move from $k=3$ to $k=5$. However, as the RF+LIME considers each time instance of a trip independently, its performance degrades. It captures the dominant features responsible for the driving behavior change within the current time window, contrary to inspecting past time windows' impact.

To have a glimpse, we present the explanatory features ($\mathcal{F}_{GEN}$) vs. human-annotated ones ($\mathcal{F}_{GT}$) in Table~\ref{table:eval} for a sample of three test instances where the similarity (Dice coefficient) is comparatively lower. Interestingly, when there is a mismatch, we observe that the corresponding features from the model-generated and human-annotated ones are conceptually related for most of the time. Additionally, a positive high mean ATE value for the model-generated mismatched features signifies that the model perceived those features as more causal than normal human perception. It can be noted that an ATE value $\geq 1$ indicates high causal relationships between the features and the corresponding effect (changes in the driving behavior). For example, in test instance $\#2$, the mismatched features are \textit{Sideslip} (for human generated) and \textit{Congestion} (for model generated), where \textit{Congestion} was relatively more causal, affecting the change in the driving behavior. By manually analyzing this instance and interviewing the corresponding driver, we found that he indeed made a minor sideslip on a congested road. Indeed, the driver was not very comfortable in driving a manually-geared car on a congested road.

\begin{figure}[!ht]
    \centering
    \includegraphics[width=0.8\columnwidth,keepaspectratio]{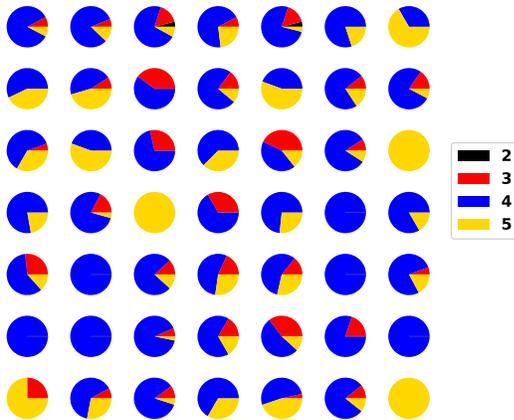}
    \caption{Generated Map from SOM for a $7\times7$ Network (Scaled Down)}
    \label{fig:model-trainsom}
\end{figure}

\subsection{Ablation Study}\label{section:ablation}
Next, we understand the importance of different feature categories corresponding to the driving maneuvers and on-road spatial events, as described in \S\ref{problem-statement}, on the overall performance of \ourmethod. To study the impact of driving maneuvers and spatial features, we implement SOM, excluding each of the above feature classes one at a time, and evaluate $\mathcal{N}$ to inspect the importance of each. The two variants other than \ourmethod{} are constructed in the following way. \textbf{(a) \ourmethod{}-man.:} Here, we exclude the driving maneuvers $\mathcal{F}_M$ and keep $\mathcal{F}_S$ only. \textbf{(b) \ourmethod{}-spat.:} Next, we exclude the spatial features $\mathcal{F}_S$ and keep $\mathcal{F}_M$ only. We evaluate these two variants over both top-3 and top-5 generated features, along with \ourmethod{} containing all the features, as depicted in Fig.~\ref{eval}(b).
On excluding the driving maneuvers and spatial features, performance drops to $45$\% and $31$\%, respectively, for top-$5$ features. This drastic drop signifies the crucial importance of spatial features, as these are the frequently changing features responsible for fluctuating driving behavior.

\subsection{Model Insight}
To understand how the spatiotemporal dependency among different features corresponding to the driving maneuvers and various on-road spatial \micro{} are derived, we use $49$ neurons spread over a $7 \times 7$ two-dimensional array (a smaller variant of the SOM network originally used to develop the model, as the original model having $147$ neurons is difficult to visualize), fitted over $200$ trips. This instance produces a \textit{Map} as depicted in Fig.~\ref{fig:model-trainsom}, where all the given trips are assigned to each of the neurons. The scores $\mathcal{R}^P_\mathcal{U}$ are used only for visual depiction purpose of how the trips are located on the \textbf{Map}. Each trip captures the change in the driving behavior using the feature variation. The neurons with multi-color are of more importance than the mono-color, as in those, the score fluctuations are most observed. During a stand-alone trip, the features corresponding to each instance of the trip will have a similar value until there is a change in the driving behavior, thus getting assigned to the same neuron (mono-color). However, the difference in the driving behavior induces distinct feature values than the previous instances; thus, it gets assigned to a different neuron in the \textit{Map}. The neurons having multi-color, as depicted in Fig.~\ref{fig:model-trainsom}, map the trip instances where a sudden change of driving behavior has occurred.

\subsection{Dissecting \ourmethod{}}
We next benchmark the resource consumption behavior of \ourmethod, followed by an analysis of the model's significance and sensitivity. 
\begin{figure*}[!ht]
    \begin{subfigure}{0.4\columnwidth}
    \includegraphics[width=\linewidth,keepaspectratio]{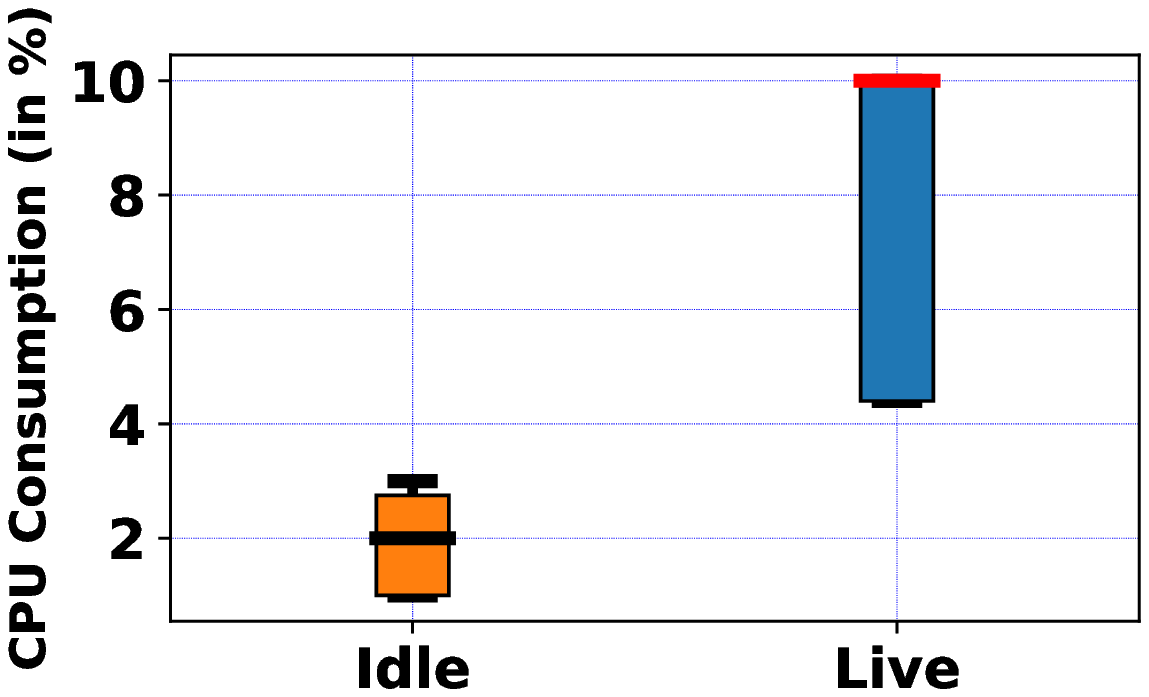}
    \caption{}
    \end{subfigure}
    \begin{subfigure}{0.4\columnwidth}
    \includegraphics[width=\linewidth,keepaspectratio]{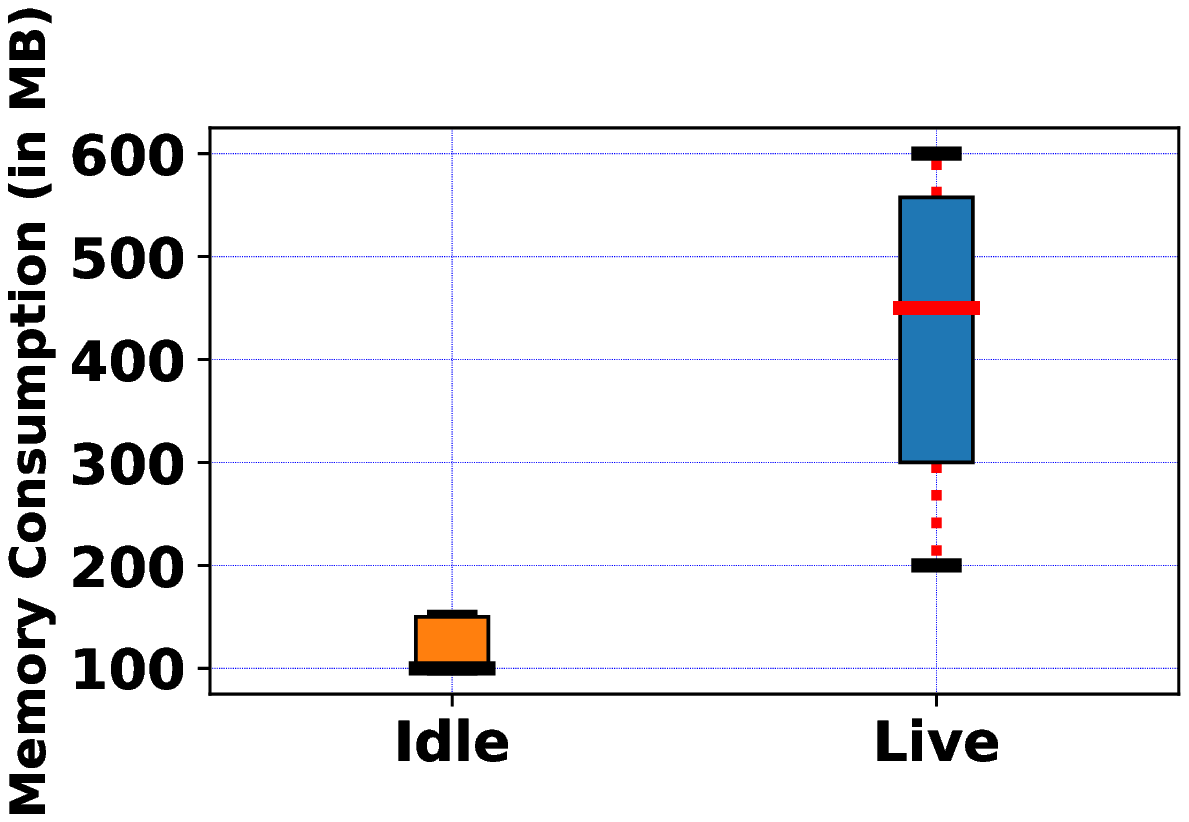}
    \caption{}
    \end{subfigure}
    \begin{subfigure}{0.4\columnwidth}
    \includegraphics[width=\linewidth,keepaspectratio]{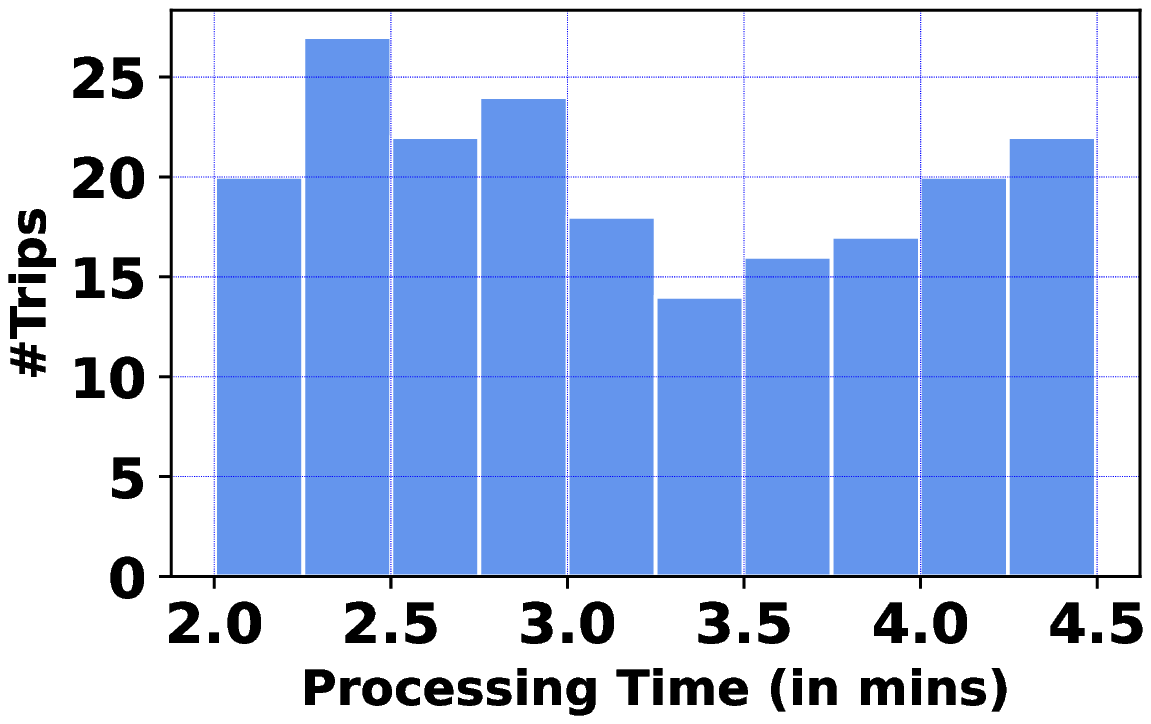}
    \caption{}
    \end{subfigure}
    \begin{subfigure}{0.4\columnwidth}
    \includegraphics[width=\linewidth, keepaspectratio]{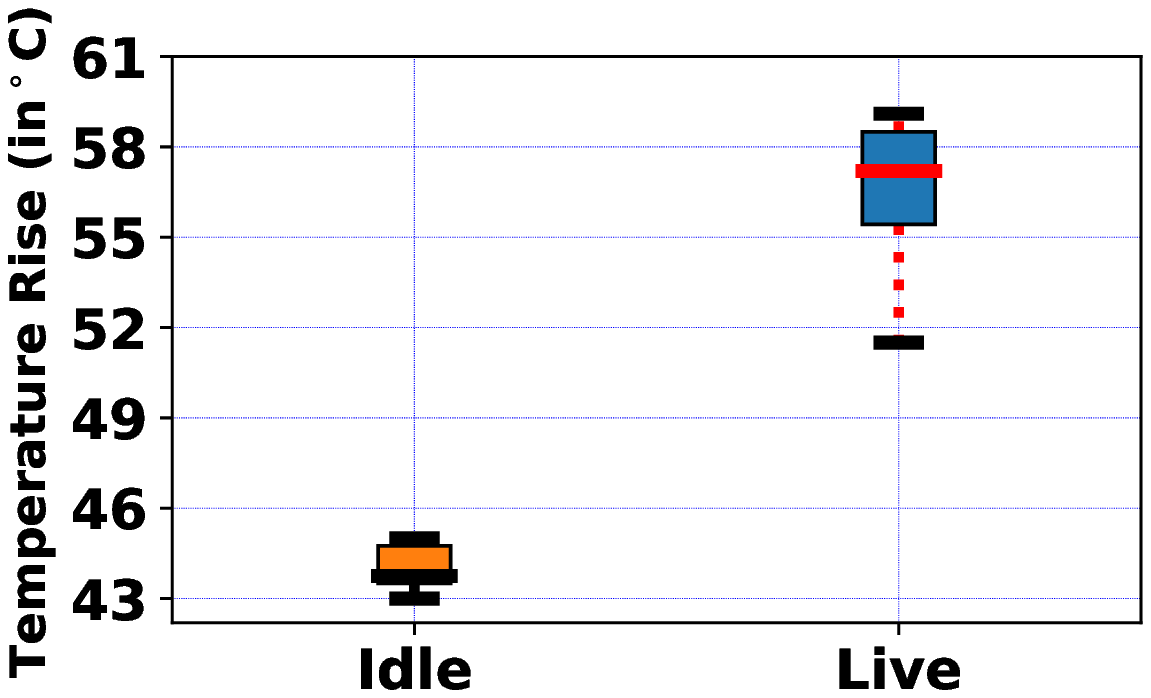}
    \caption{}
    \end{subfigure}
    \begin{subfigure}{0.4\columnwidth}
    \includegraphics[width=\linewidth,keepaspectratio]{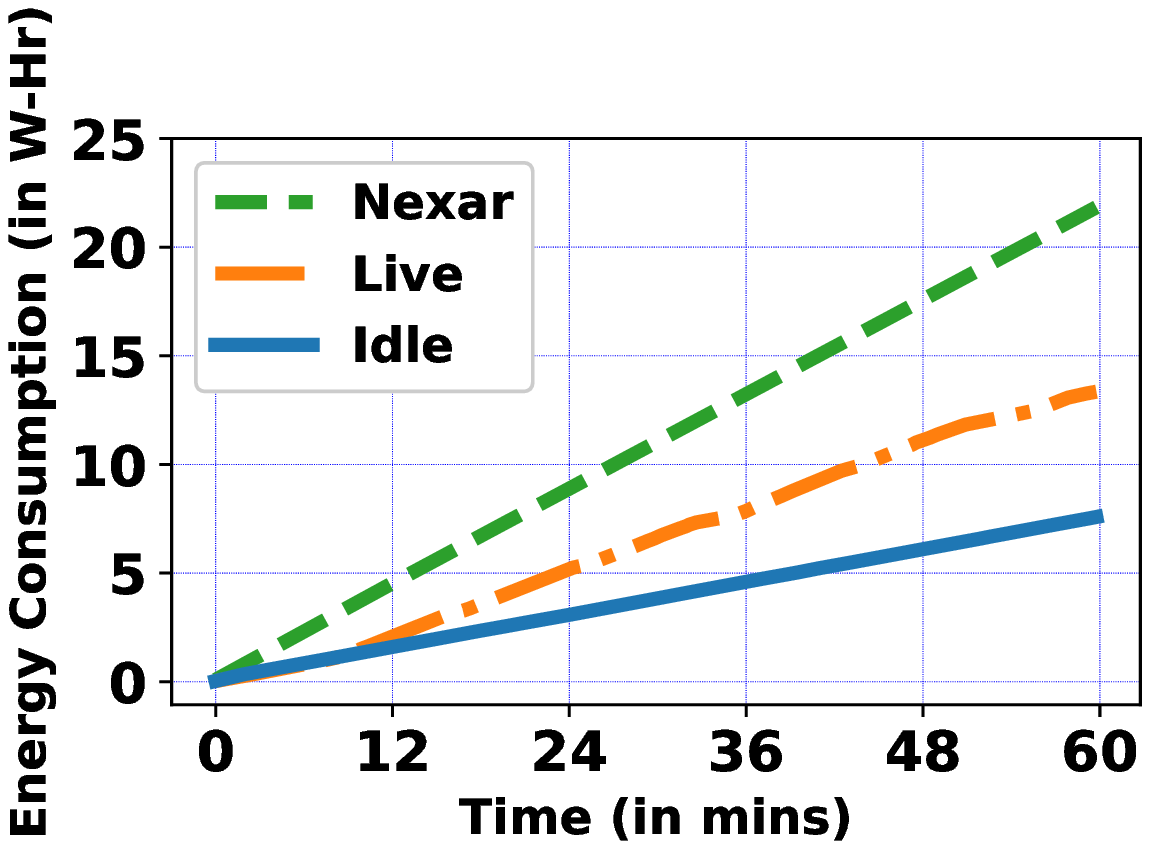}
    \caption{}
    \end{subfigure}
    \caption{Resource Consumption over the Edge-device (a) CPU Usage (b) Memory Usage  (c) Histogram of Processing Time w.r.t., \#Trips (d) Temperature Rise, (e) Energy Consumed}
    \label{fig:edge-device}
\end{figure*}

\subsubsection{\textbf{Edge-device Resource Consumption}}
We benchmark the CPU \& memory usage, processing time, temperature rise, and energy consumption over two cases: when (a) the device is idle, \& (b) \ourmethod{} is running. From Fig.~\ref{fig:edge-device}(a), we observe that in idle mode, on average, $2$\% of CPU (using ``top'' command) is used. In contrary, running \ourmethod{} acquires at most $10$\% of the processor, which is acceptable. However, the memory usage is a bit high ($\approx500$MB) mainly due to video processing overhead as depicted in Fig.~\ref{fig:edge-device}(b). Next, we show the required processing time starting from data acquisition to output generation on a number of trip basis. \ourmethod{} generates the output within $\approx 3$ minutes only for majority of the trips, further validating shorter response time (see Fig.~\ref{fig:edge-device}(c)). To further delve deeper, we also log the temperature hike (from ``vcgencmd measure\_temp'' command) and total energy consumption using Monsoon High Voltage Power Monitor~\cite{monsoon} while running \ourmethod. From Fig.~\ref{fig:edge-device}(d) \& (e), we observe that the temperature hiked at most to $59$\textdegree C, while on average, $13$ Watt-hour energy is consumed, which is nominal for any live system. To benchmark \ourmethod{}, we have also measured the energy consumption of the Nexar dashcam, which consumes $22$ Watt-hour on an average, while capturing very few driving maneuvers (say, hard brake) without any context. This further justifies that \ourmethod{} never exhausts the resources on the edge-device and is can accurately detect the \micro{} precisely.

\begin{figure}[]
\centering
\begin{subfigure}{0.3\columnwidth}
    \includegraphics[width=\linewidth,keepaspectratio]{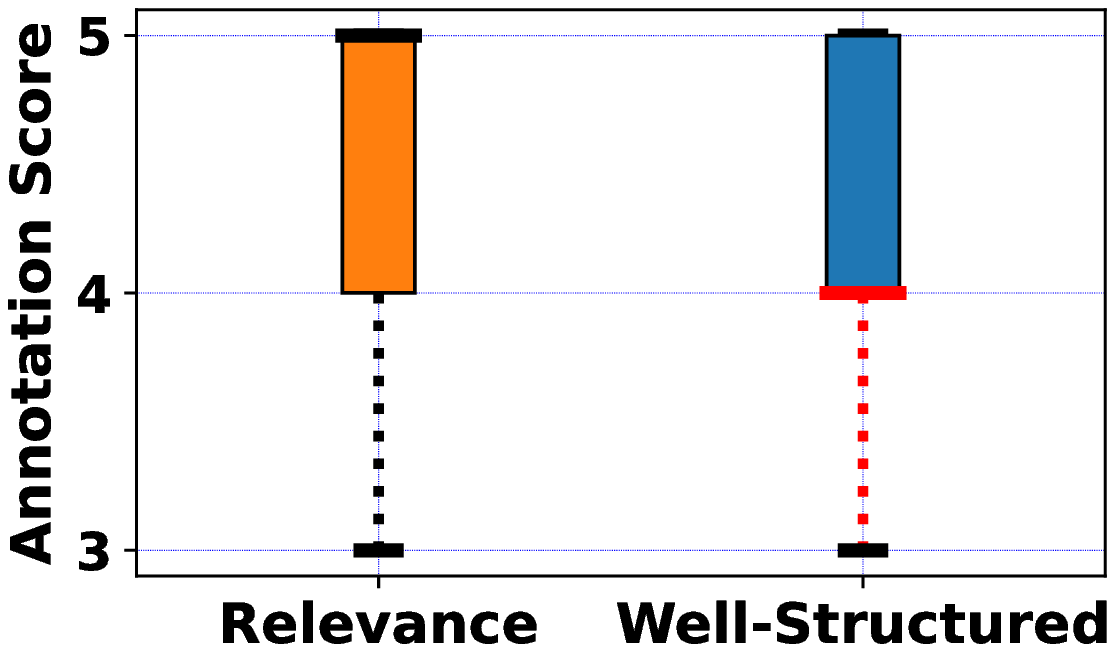}
    \caption{}
    \end{subfigure}
    \begin{subfigure}{0.3\columnwidth}
    \includegraphics[width=\linewidth,keepaspectratio]{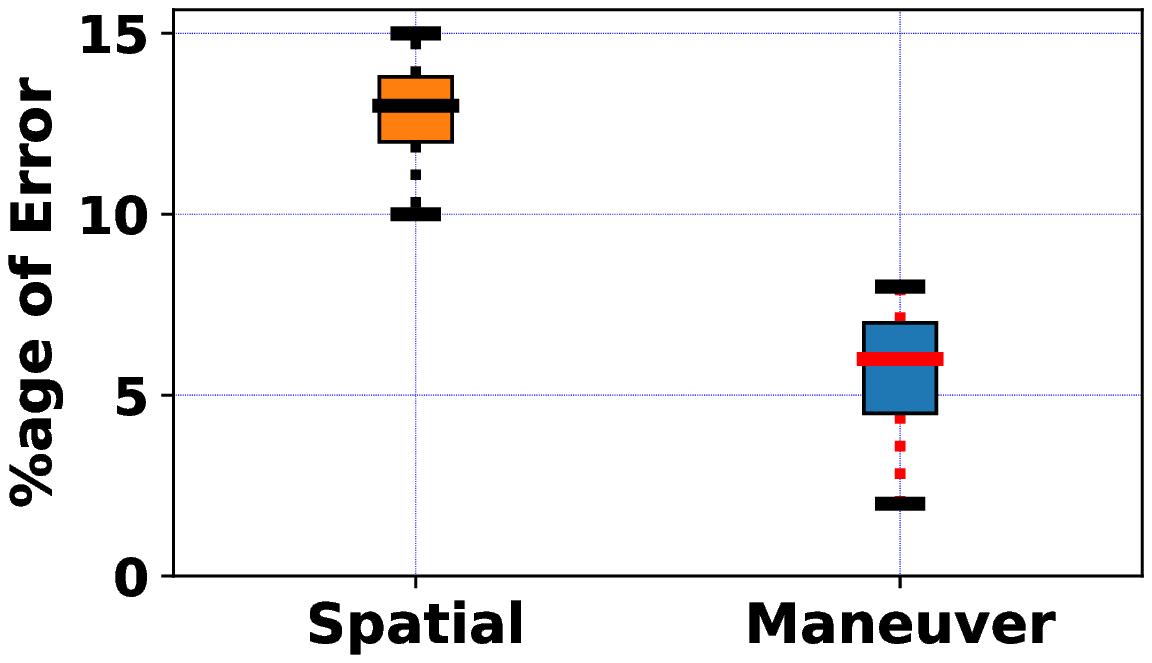}
    \caption{}
    \end{subfigure}
    \begin{subfigure}{0.3\columnwidth}
    \includegraphics[width=\linewidth,keepaspectratio]{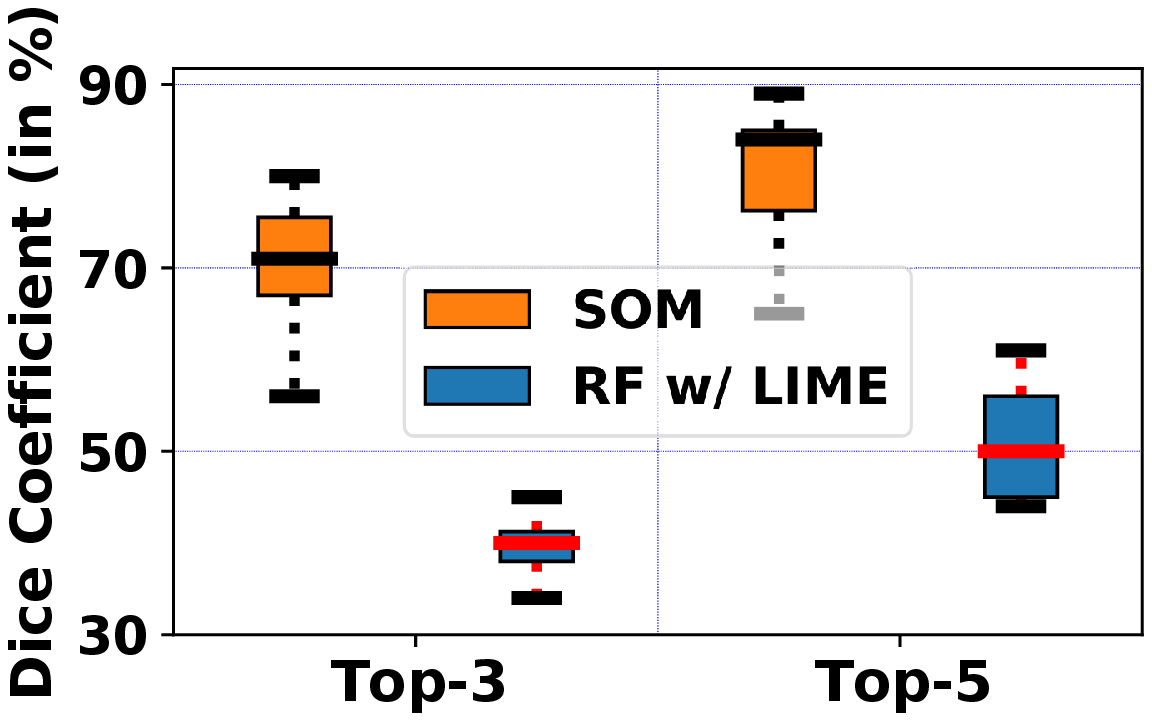}
    \caption{}
    \end{subfigure}
\caption{(a) Significance of \ourmethod{} (b) Sensitivity Analysis of \ourmethod{} (c) Performance on BDD Dataset}
\label{dissect}
\end{figure}

\subsubsection{\textbf{Significance of Generated Explanation}}
Next, we check how significant our generated explanations are. As reported in \S\ref{annotate-micro}, we plot the distribution of annotated scores (given by the recruited annotators) for the two fields -- ``\textit{Relevance}'' and ``\textit{Well-Structured}''. ``Relevance'' signifies the generated explanation's applicability in explaining unexpected events. In contrast, ``Well-structured'' indicates how well interpretative the generated sentences are as per human cognition. Fig.~\ref{dissect}(a) depicts a median value of $5$ and $4$ for ``Relevance'' and ``Well-Structured'', respectively, which further justifies the credibility of \ourmethod. We also compute the similarity between the human-annotated and model-generated sentences and obtain a minimum, maximum, and mean similarity value as $51.33$\%, $85.5$\% \& $70.57$\%, respectively, using the TF-IDF vectorizer. Thus \ourmethod{} resembles human cognition level up to an indistinguishable level (between a human and model) of auto-generating a contextual explanation, which further shows its applicability to give feedback to the stakeholders for their decision-making procedure.

\subsubsection{\textbf{Sensitivity of \ourmethod{}}}
Finally, we inspect the micro-events that \ourmethod{} fails to capture. Because, apart from a model's efficiency, we must also look into its deficiency to analyze how much that might affect the overall performance. Especially, this is important in the case where stakeholders are boosting/penalizing the driver's profile. As depicted in Fig.~\ref{dissect}(b), incompetence to capture both the spatial and maneuvers is low. Although this might lead to degraded model performance, as studied in \S\ref{section:ablation}; driving maneuvers ($\mathcal{F}_M$) do not contribute superiorly to model performance due to the inter-dependency on spatial features ($\mathcal{F}_S$). But for $\mathcal{F}_S$, the \textbf{Percentage of Error} is still $\leq13$\%, making the system less sensitive into generating error-prone contextual explanations.

\subsection{Offline Performance}
Finally, we report the accuracy of our system over the BDD dataset comprising $17$ hours of driving data over $1.5k$ trips using $\mathcal{N}$. As depicted in Fig.~\ref{dissect}(c), \ourmethod{} performs quite well on pre-recorded data, with $\mathcal{N}=\{71\%,84\%\}$, for top-3 and top-5 features. We observe that SOM can identify the \micro{} in a better way for offline analysis with a public dataset. However, as running the system live is essential for a realistic driving environment other than offline analysis, this much of slight accuracy drop can be endured.

%% file: sections/Conclusion.tex
\section{Conclusion}\label{conclusion}
This paper developed an intelligent system on the edge-device called \ourmethod{} leveraging multi-modalities to detect the \micro{} responsible for unexpected fluctuations in driving behavior. The human-interpretable explanations generated by \ourmethod{} show their relevance and credibility in identifying such context. Further, the spatiotemporal dependency among various features is inspected in an unsupervised manner to capture a diverse set of driving scenarios. Additionally, the resource-friendly deployment over a live testbed further validates \ourmethod{}. Although our study captures the context where each feature's contribution is taken independently, inter-feature dependency is not captured explicitly. For instance, say, a driver suddenly weaves while taking a turn to avoid colliding with a crossing pedestrian, making the following vehicle's driver slam the brake. Here, the first driver's action is due to the crossing pedestrian, which in turn impacts the second driver's action. The analysis of such complex and collective interactions among the vehicles needs a more sophisticated system, possibly a different modality that can connect the inter-vehicle interactions. However, \ourmethod{} provides a simple, in-the-silo solution that can be independently deployed over vehicles with a dashboard-mounted edge-device or dashcam.
